\documentclass[twocolumn]{aastex62}

\accepted{4/20/2020}
\submitjournal{ApJ}

\shorttitle{Cluster UV Background}
\shortauthors{Welch et al.}

\begin{document}

\title{Galaxy Cluster Contribution to the Diffuse Extragalactic Ultraviolet Background}

\correspondingauthor{Brian Welch}
\email{bwelch7@jhu.edu}

\author{Brian Welch}
\affiliation{Department of Physics and Astronomy, The Johns Hopkins University 
3400 N Charles St. 
Baltimore, MD 21218}

\author{Stephan McCandliss}
\affiliation{Department of Physics and Astronomy, The Johns Hopkins University 
3400 N Charles St. 
Baltimore, MD 21218}

\author{Dan Coe}
\affiliation{Space Telescope Science Institute 
3700 San Martin Drive 
Baltimore, MD 21218}

\begin{abstract}

The diffuse ultraviolet background radiation has been mapped over most of the sky with 2\arcmin \ resolution using data from the \textit{GALEX} survey. We utilize this map to study the correlation between the UV background and clusters of galaxies discovered via the Sunyaev-Zeldovich effect in the \textit{Planck} survey. We use only high Galactic latitude ($|b| > 60 \degr $) galaxy clusters to avoid contamination by Galactic foregrounds, and we only analyze clusters with a measured redshift. This leaves us with a sample of 142 clusters over the redshift range $0.02 \leq z \leq 0.72$, which we further subdivide into four redshift bins. In analysing our stacked samples binned by redshift, we find evidence for a central excess of UV background light compared to local backgrounds for clusters with $z<0.3$. We then stacked these $z<0.3$ clusters to find a statistically significant excess of $12 \pm 2.3$ photon cm$^{-2}$ s${-1}$ sr$^{-1}$ \AA $^{-1}$  \ over the median of $\sim 380$ photon cm$^{-2}$ s${-1}$ sr$^{-1}$ \AA $^{-1}$  \ measured around random blank fields. We measure the stacked radial profile of these clusters, and find that the excess UV radiation decays to the level of the background at a radius of $\sim 1$ Mpc, roughly consistent with the maximum radial extent of the clusters. Analysis of possible physical processes contributing to the excess UV brightness indicates that non-thermal emission from relativistic electrons in the intracluster medium and faint, unresolved UV emission from cluster member galaxies and intracluster light are likely the dominant contributors.

\end{abstract}


\section{Introduction}

Diffuse background radiation can be observed across the entire electromagnetic spectrum, from the Cosmic Microwave Background (CMB) radiation \citep{cmb} to the highest energy gamma rays \citep{gammaray}. In the ultraviolet (UV), the largest contributor to the background is of Galactic origin. Previous studies have seen clear correlations between diffuse UV excesses and tracers of interstellar gas and dust, indicating that the primary contributor is scattered starlight \citep{bowyer91, murthy2010}. However, there is a notable non-scattered component to this background, observable as a $\sim 300$ photon cm$^{-2}$ s${-1}$ sr$^{-1}$ \AA $^{-1}$  \  (continuum units, CU) background observed in low column density regions near the Galactic poles \citep{henry91, hamden13, murthy2016}. While there is some debate as to the origin of this portion of the background, some component is likely extragalactic. \cite{murthy2016} calculated that of the $ \sim 300 $ CU background observed at the Galactic poles, around 100 CU is unexplained by galactic processes, and is therefore likely extragalactic in origin. Recently, \cite{chiang19} used a combination of \textit{GALEX} and SDSS data to calculate that the total extragalactic background of $89^{+28}_{-16}$ CU, while \cite{Akshaya_2018} used \textit{GALEX} data at the Galactic poles to calculate a total extragalactic background of $114 \pm 18$ CU. 

The sources of extragalactic ultraviolet background radiation are generally assumed to be active galactic nuclei and star-forming galaxies \citep{upton-sanderbeck,becker}. These appear to be the dominant sources of metagalactic ultraviolet radiation, and they are believed to be the primary sources of ionizing ultraviolet radiation in the Epoch of Reionization. However, precise measurements of the UV background are difficult, and recent measurements of the relative contribution of star-forming galaxies and AGN indicate there may be other extragalactic sources. Using number counts of galaxies detected in the FUV band of the \textit{GALEX} survey, \cite{xu05} calculated a total contribution from galaxies (in units of $\lambda F_{\lambda}$) of $1.03 \pm 0.15 \text{ nW m}^{-2} \text{ sr}^{-1} $, or $ 51.5 \pm 7.5$ CU. \cite{Voyer11} performed a similar calculation with HST data. They used number counts of field galaxies from the GOODS fields, the Deep Field North, and the Ultra-Deep Field with FUV magnitudes between 21 and 29 AB in the ACS Solar Blind Channel, and calculated a contribution to the UV background of 65.9 to 82.6 CU. More recently, \cite{chiang19} used a broadband intensity tomography method with a combination of \textit{GALEX} and SDSS data to calculate the combined contribution from galaxies and AGN to be $73 \pm 8$ CU. While it is clear from these measurements that star-forming galaxies and AGN are the dominant contributors to the extragalactic background, they each leave room for additional sources. Additionally, \cite{Akshaya_2018,Akshaya_2019} used \textit{GALEX} data to tabulate the contributions to the UV background from dust-scattered starlight and known extragalactic sources near the Galactic poles. They found an unexplained offset of $\sim 200$ CU at zero dust column density ($E(B-V) = 0$) in the FUV band. While they do not identify the source of this offset, they speculate that hitherto unknown extragalactic sources could be contributing to this offset. 

In this paper, we explore another possible source contributing to the diffuse ultraviolet background: massive clusters of galaxies. While not traditionally associated with high ultraviolet luminosities, the high density of galaxies and hot gas in massive galaxy clusters suggests them as candidate sources of excess diffuse light. Utilizing UV background data from the \textit{GALEX} survey \citep{murthy14} and massive galaxy clusters from the \textit{Planck} survey \citep{psz}, we measure the correlation between galaxy clusters and diffuse UV background light. 

This paper is organized as follows. Section 2 describes the data from the \textit{GALEX} and \textit{Planck} surveys. Section 3 details the analysis methods used to investigate the correlation between clusters of galaxies and diffuse UV background light. Section 4 presents and discusses the results of our analysis, and Section 5 presents our summary and conclusions.

\section{Data}

\subsection{Galex FUV Background Catalogs}

\textit{GALEX} was an orbiting ultraviolet observatory that made use of a 50 cm Ritchey-Chretien telescope to image a $1.2\degr$ circular field of view onto two detectors \citep{martin05}. The detectors operate in two different bandpasses, one in the far-UV from 1350 - 1750 \AA, and the other in the near-UV from 1750 - 2750 \AA. 

Our analysis focuses on the FUV bandpass data from the GR6/GR7 data release, which was further processed by \cite{murthy14} into a map of diffuse background FUV flux. Briefly, this map was made by masking all point sources found in the standard \textit{GALEX} data reduction pipeline \citep{morrissey07} in the raw images, then binning the image data into 2\arcmin \  pixels. Masked pixels were ignored in the binning, thus replacing them with an average from the full 2\arcmin \  binned pixel. 

We restrict our analysis to Galactic latitudes $|b|>60\degr$. This allows us to avoid the most significant Galactic contributions to the diffuse UV background light, and focus on the extragalactic component. Restricting our sample this way ensures that our results are not significantly biased by Galactic scattered stellar light. 

\subsection{Planck SZ Cluster Sample}

For this work, we analyze a subsample of clusters from the second \textit{Planck} catalog of Sunyaev-Zeldovich sources \citep{psz}. These clusters are detected via their thermal Sunyaev-Zeldovich effect signal \citep{sz_effect}, wherein hot gas in the intracluster medium inverse Compton scatters photons from the CMB to higher energies. This effect causes a decreased intensity at lower frequencies and an increased intensity at higher frequencies. 

\begin{figure}
    \centering
    \includegraphics[width=0.48\textwidth]{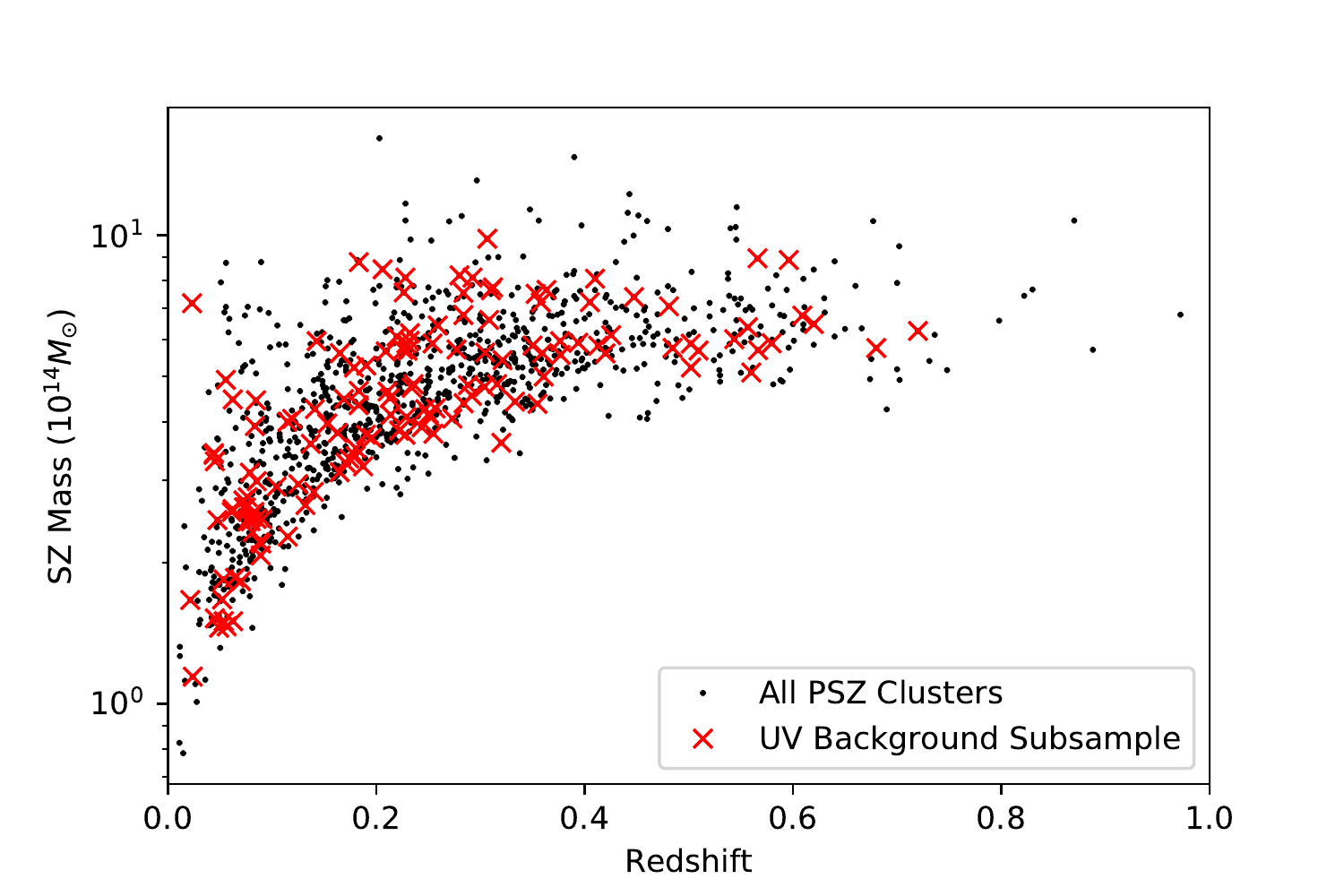}
    \caption{Distribution of \textit{Planck} SZ cluster masses  (in units of $10^{14} M_{\odot}$) and redshifts. Clusters within our sample are highlighted with a red X.}
    \label{psz_mass}
\end{figure}

For our investigations, we select only clusters which have a measured redshift and SZ mass. We further select only clusters which lie in the Galactic cap region, with latitudes $|b|>60 \degr$. We then visually inspected the sample and discarded any clusters with incomplete or irregular \textit{GALEX} background data, for example those near large holes in the background map or with partial data near the edges of our window.  This left a sample of 185 clusters. A final cut was applied after matching the cluster positions to the \textit{GALEX} data. Clusters with no UV background data present within the central region, as described in Section \ref{section:methods}, were excluded from our analysis. This left a final sample of 142 clusters. The masses and redshifts of these clusters are shown in Figure \ref{psz_mass}.

\begin{figure}
    \centering
    \includegraphics[width=0.4\textwidth,trim={1.5cm 1cm 1cm 1cm},clip]{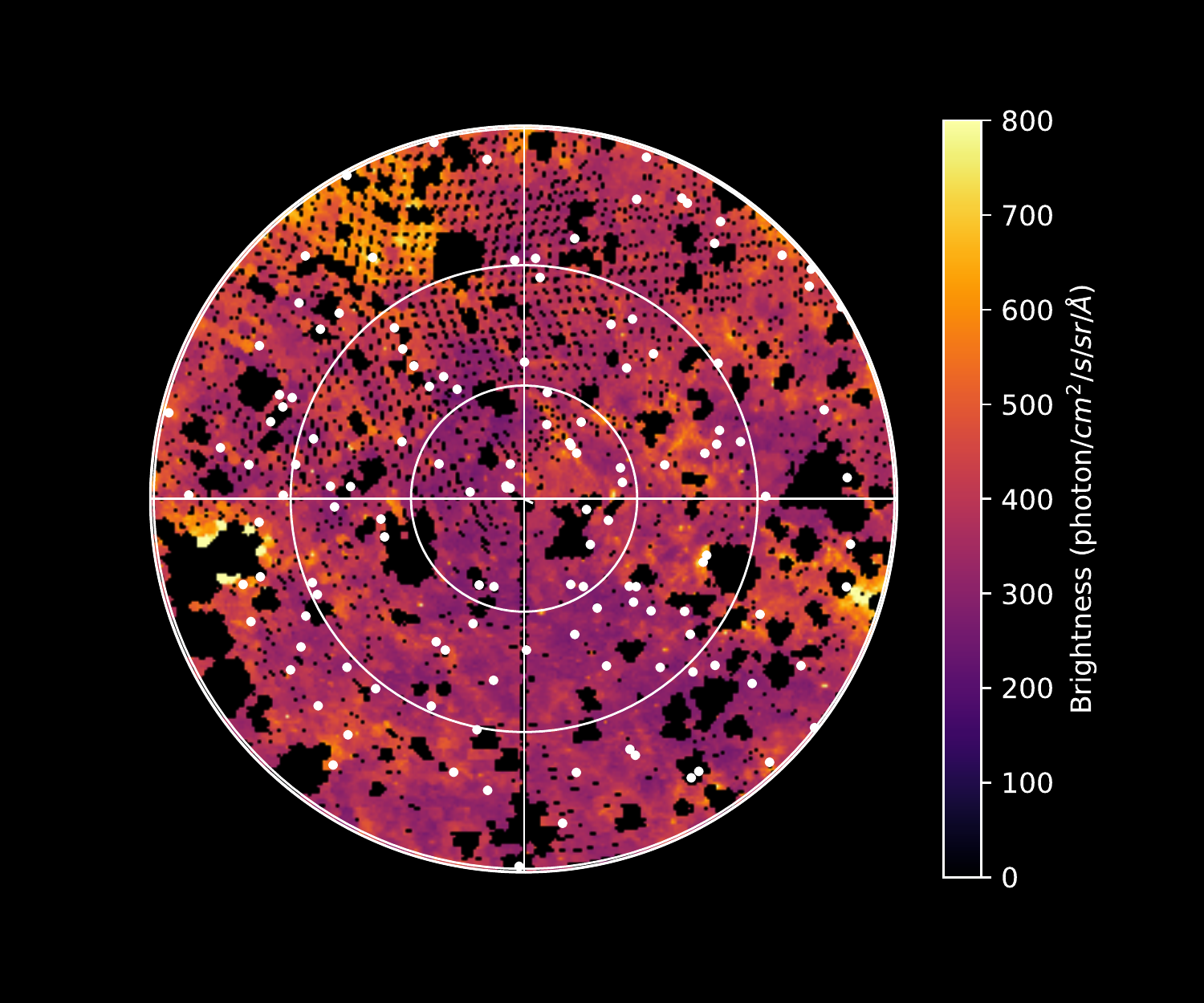}
    \includegraphics[width=0.4\textwidth,trim={1.5cm 1cm 1cm 1cm},clip]{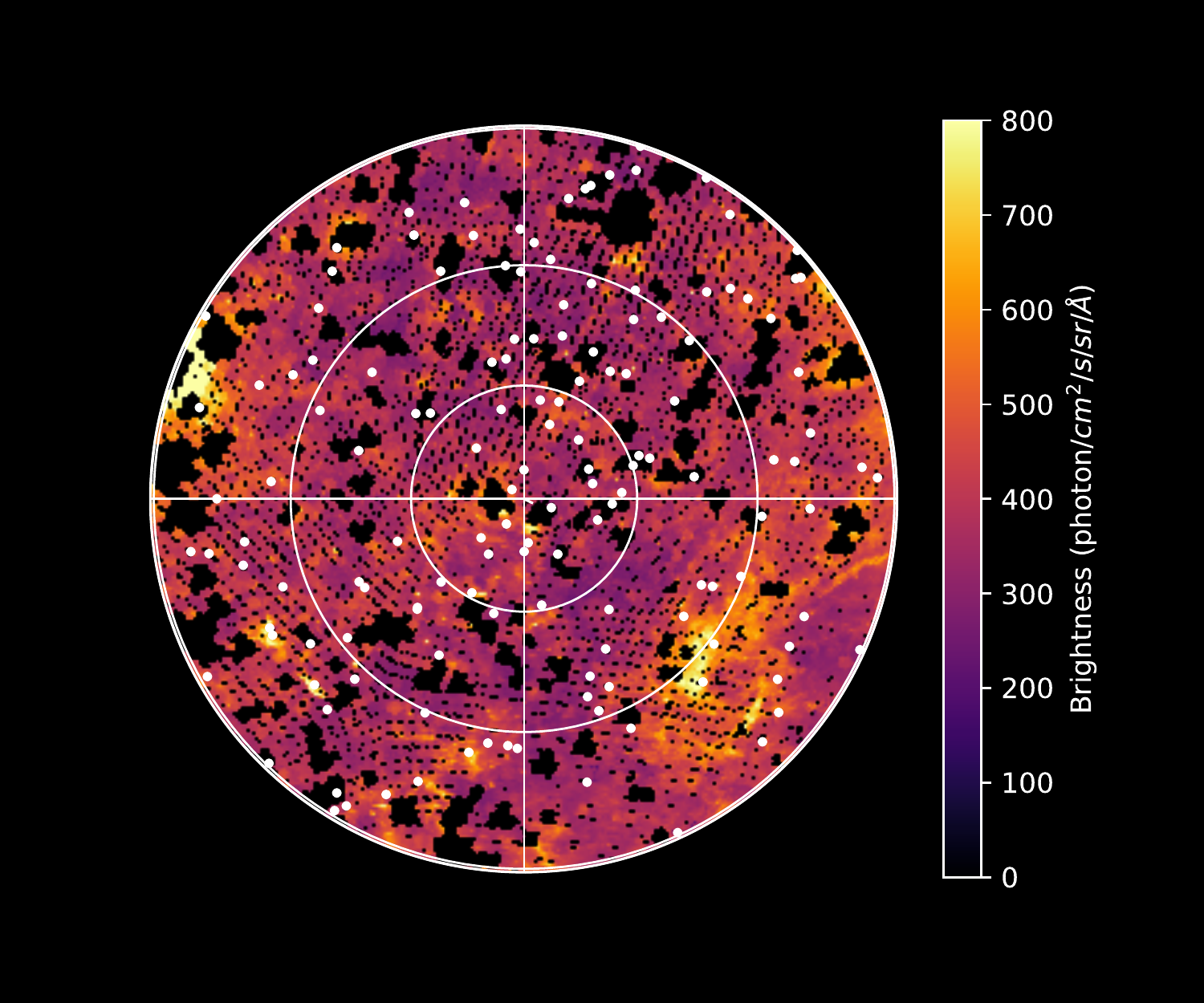}
    \caption{Northern (left) and southern (right) Galactic cap \textit{GALEX} FUV background map, down to $|b| = 60\degr$. Locations of the \textit{Planck} clusters used in this study are marked with white circles.}
    \label{polarmap}
\end{figure}

\section{Methods}
\label{section:methods}

To analyze the level of UV background light around our sample of galaxy clusters, we created radial UV brightness profiles. These were made by matching all \textit{GALEX} data within 10 Mpc of the cluster centers, binned into annuli sized to accommodate the 2\arcmin \  background map resolution. Any clusters with no \textit{GALEX} data matched within the central region are removed from the final sample to ensure all fully analyzed clusters have complete data. To overcome the high variability between individual clusters, we stacked our sample to create a median cluster radial brightness profile. Uncertainties on this profile were estimated by randomly resampling UV background fluxes with replacement at each radius. \cite{murthy14} calculates a photon count uncertainty of 18 CU per 2 \arcmin \  pixel. To account for this uncertainty, we add Gaussian noise with standard deviation $\sigma = 18 \text{CU} / \sqrt{N_{pix}}$ to each sample drawn, where $N_{pix}$ is the number of 2\arcmin \  pixels included in the given annular bin. We used the standard deviation of the mean of 1000 random resampling iterations as the flux uncertainty at a given radius. 

\begin{table}
\centering
\begin{tabular}{ccc} 
\hline 
Redshift Bin & Number of Clusters & Annulus Size (Mpc) \\
\hline
$0.02 < z \leq 0.1$ & 37 & 0.25 \\
$0.1 < z \leq 0.2$ & 27 & 0.4 \\ 
$0.2 < z \leq 0.3$ & 35 & 0.5 \\ 
$z>0.3$ & 43 & 1.0 \\ 
\hline 
\end{tabular} 
\caption{Summary of our cluster sample binned by redshift.}
\label{table:zbin}
\end{table}

We divided our cluster sample into four redshift bins to maximize the resolution of our radial profile measurements at lower redshifts while still being able to analyze the full redshift range. The redshift bins and associated annular sizes are summarized in Table \ref{table:zbin}. The annulus sizes were chosen to match the 2\arcmin \ pixel scale at the high end of each redshift bin, thus ensuring that at least one pixel will be included within the central region. 

We found evidence of a central excess in each of our redshift bins below $z=0.3$, discussed further in Section \ref{section:results}. This led us to stack our full sample of $z \leq 0.3$ clusters in an attempt to maximize our signal-to-noise, giving us a sample of 99 clusters. We stacked these clusters with annular bin sizes of 0.5 Mpc, corresponding to the maximum spatial resolution possible at $z=0.3$. 

The shape of the stacked cluster profile is informative on its own; However comparison with a blank field profile provides a useful context for this measurement. The blank field profile is constructed by randomly selecting 185 locations on the sky, subject to the same $|b|>60\degr$ restriction as our cluster sample.  These locations are matched to the \textit{GALEX} data in the same way as the cluster sample. The same cut on points with no central data are applied, leaving a final random field sample of $\sim 120 - 150$ points per iteration, similar in number to our final cut cluster sample of 142 clusters. Blank fields obviously do not have an associated redshift, so we randomly select a redshift for each field within the range $[0.01,0.3]$, encompassing the redshift range of our primary stacked sample. A stacked radial profile is then computed for this set of random blank fields, with the same stacking procedure as in the cluster sample. To account for cosmic variance and variations in the UV background radiation field, we repeat this random field measurement 1000 times. The uncertainty values for the blank field radial profile measurements are calculated as the standard deviation of the mean of the 1000 iterations, mimicking the randomly resampled uncertainty calculation of the cluster sample.

\section{Results and Discussion}
\label{section:results}

\begin{figure}
\centering
\includegraphics[width=0.5\textwidth]{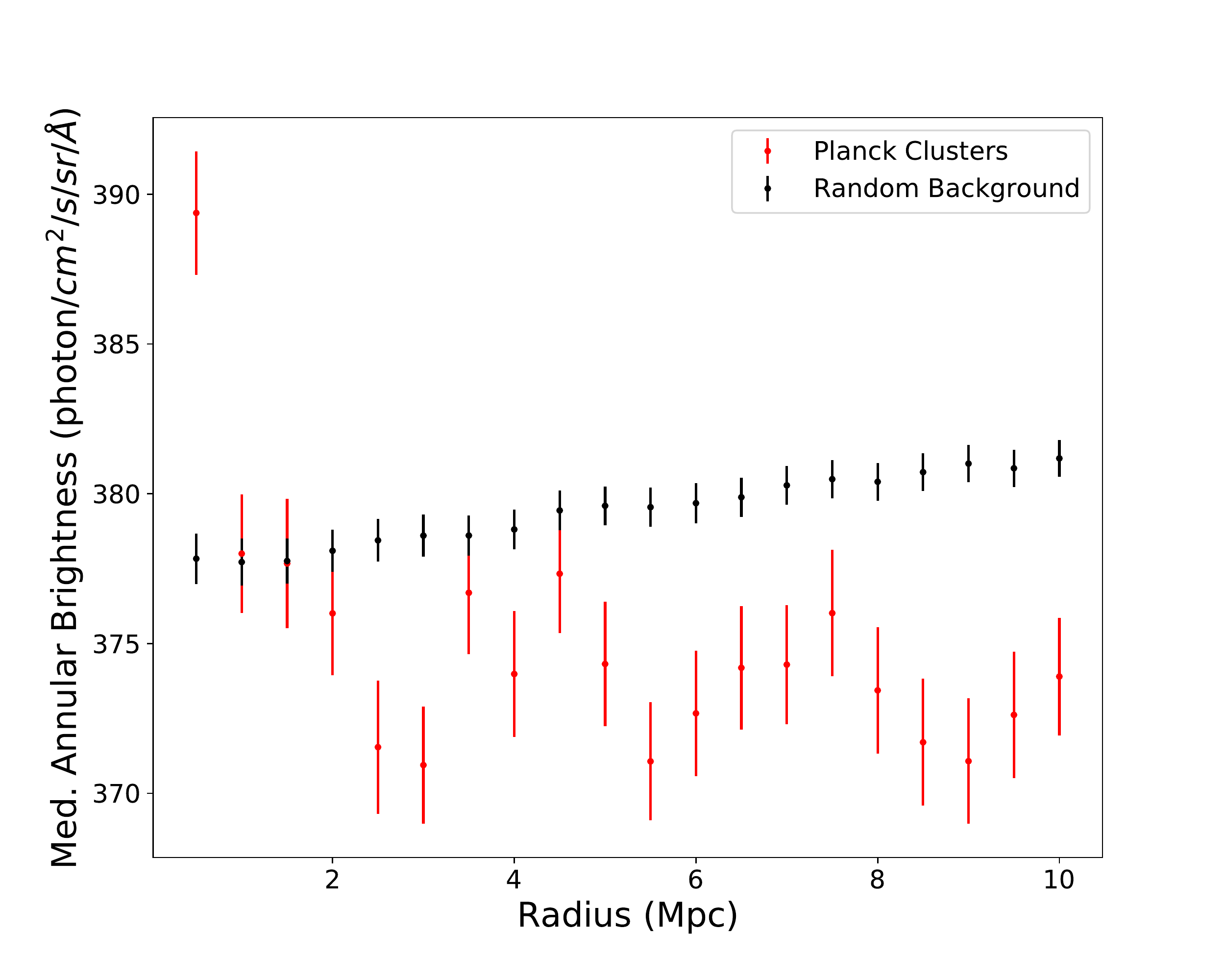}
\caption{FUV background fluxes measured around $z \leq 0.3$ Planck SZ galaxy clusters (red points with errorbars) and around randomly selected points on the sky (black points with errorbars). Uncertainty calculations are discussed in detail in the text. The excess flux around galaxy clusters is clearly visible at $r= 0.5$ Mpc, and rapidly falls off to the background level beyond that point.}
\label{rad_prof}
\end{figure}

Our stacked radial profile of all clusters with $z \leq 0.3$ (Figure \ref{rad_prof}) shows a clear excess of UV background radiation at the locations of Planck SZ-selected clusters. The peak excess of $12 \pm 2.3$ CU (a $5.0 \sigma$ detection) is coincident with the cluster center. 

The cluster signal decays rapidly further from the cluster center. The UV background brightness in the cluster fields drops back to be consistent with the random fields within 1 Mpc, roughly consistent with the virial radii of massive clusters. The coarse binning of the UV background map prevents more detailed analysis of the shape of the brightness profile for the full $z \leq 0.3$ sample. 

\begin{figure}
    \centering
    \includegraphics[width=0.5 \textwidth]{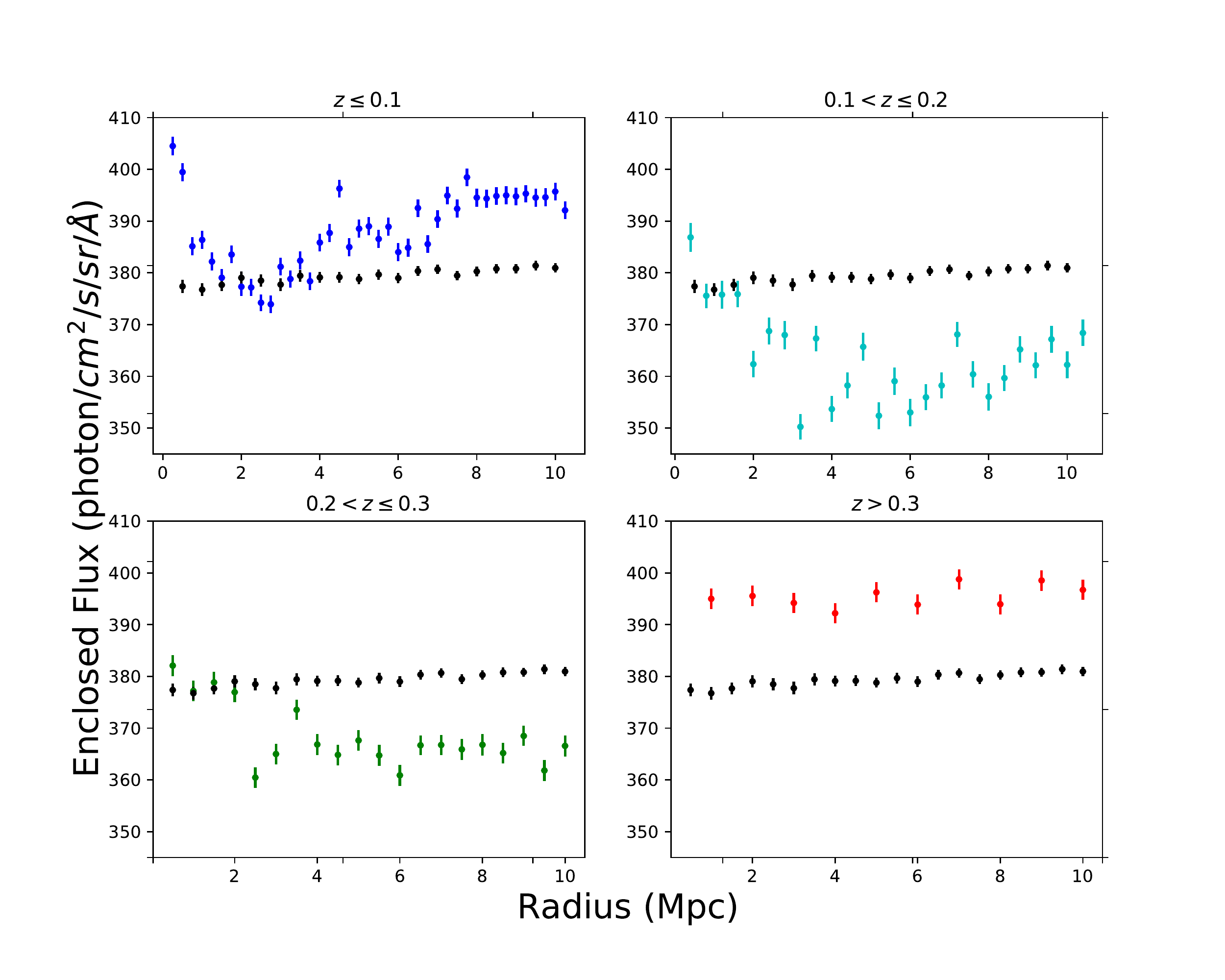}
    \caption{UV background radial profiles in each of four redshift bins. Black points in each plot are blank field measurements, as described in the text. Central excesses are visible in each bin below $z=0.3$.}
    \label{zbin_profile}
\end{figure}

Dividing the sample into smaller redshift bins allows us to look in more detail at the central region of nearer clusters (Figure \ref{zbin_profile}). In our lowest redshift subsample ($z \leq 0.1$) our 0.25 Mpc annuli give us the greatest chance to probe the centers of our clusters. In this sample we see that the two innermost points ($r<0.25$ Mpc and $0.25<r<0.5$ Mpc) are both elevated above the background level, indicating that the background emission source extends over the full central 0.5 Mpc. Beyond $z=0.1$, our annulus sizes become too large (0.4 Mpc for $0.1 < z \leq 0.2$) to see this innermost detail. We continue to see a clear central excess in the $0.1 < z \leq 0.2$ bin, while the $0.2 < z \leq 0.3$ redshift bin shows mild evidence of a central excess. The highest redshift bin ($z>0.3$) shows no evidence of central excess. We attribute this lack of signal in the highest redshift bin to the larger annuli required to match the background map resolution. Our lower redshift samples show that the excess has decayed to the background level within 1 Mpc, so annuli of 1 Mpc cannot detect the central peak. Additionally, \cite{murthy14} explains that the 2\arcmin \ pixels display the average of all original \textit{GALEX} pixel background values within the region. Thus any signal of central excess will be further washed out in these higher redshift clusters. 


Below we estimate contributions to the cluster excess from various sources.  

\subsection{BCG or Infalling Jellyfish Galaxies}
It is possible that this central excess is originating from the brightest cluster galaxy (BCG). Previous studies have found that cluster BCGs can be UV bright, particularly in cases of cool-core clusters where the BCG is actively forming stars \citep{hicks_coolcore}. These galaxies, with star-formation rates $0.01 \leq SFR \leq 10$, would be bright enough to provide the full excess we measure within the central 0.5 Mpc of the cluster. However, these galaxies would be detectable in the \textit{GALEX} survey, with an all-sky survey limiting magnitude of 19.9 in the FUV. Therefore these galaxies would most likely be masked in the background maps. It is possible that incomplete masking could allow UV flux from the BCG to leak into the maps and contribute to our observed signal. It is impossible to determine the exact contribution of incomplete masking on our result without remaking the background maps. 

Additionally, BCG star formation studies have focused on much smaller scales ($\sim 10$ kpc as opposed to Mpc).  Finally, our sample includes many clusters that are not known to have active star formation in their central galaxies, indicating alternative sources of UV background light production. Therefore it is unlikely that BCG star formation is the primary contributor to the excess UV flux observed.

Similarly, it is possible that star formation in infalling field galaxies or jellyfish galaxies being pressure stripped as they fall into the cluster could contribute to our signal. While these galaxies would likely have elevated star formation rates, and thus greater emission in the UV, they would likely be bright enough to be detected by GALEX. They would therefore be masked out of our background sample, and unlikely to contribute to the observed excess. 

\subsection{Cluster Gas Emission Estimates: \\ Thermal Brehmsstrahlung and Inverse Compton Scattering}

Clusters are full of hot gas, which is known to radiate brightly in X-rays through the combined effects of thermal brehmsstrahlung and inverse Compton scattering of CMB photons off hot electrons. We considered these as possible contributors to our measured signal in the FUV, as these spectra for a typical cluster continue to longer wavelengths. 

In our FUV bandpass, the brehmsstrahlung luminosity is $L_{\nu} \sim 10^{27} \ $ erg sec$^{-1}$ Hz$^{-1}$ \citep{Sarazin99,sarazin_merger}.  We assume a cluster redshift of $z = 0.3$, because most of our clusters are within this range and changes in assumed redshift do not drastically impact our results. This gives a calculated FUV surface brightness of $\sim 0.2$ CU  over an area of 2\arcmin, matching the resolution of the background map. The fact that this calculated surface brightness is much smaller than our measured excess indicates that brehmsstrahlung radiation from cluster gas is likely not a dominant contributor to our observed signal. 

Another possibility is that shocks generated by cluster mergers accelerate relativistic electrons within the cluster, which emit ultraviolet light through the inverse Compton effect. \cite{sarazin_merger} calculated that these electrons would have lifetimes of order a Hubble time, so the cluster would not have to have been through a recent merger for this effect to be visible. They also calculated a spectrum for the emission, giving a luminosity in the FUV bandpass of $L_{\nu} \sim 10^{28.5} \ $ erg sec$^{-1}$ Hz$^{-1}$. Again assuming a cluster redshift $z = 0.3$, we calculate the FUV surface brightness of this emission to be 8 CU,  again over an area of 2\arcmin.  This indicates that emission from shock-accelerated electrons in the intracluster medium is likely a significant contributor to the signal we observe.

\subsection{Cluster Emission from Stripped Stars}
\label{section:clusteremission}

While cluster galaxies generally contain old stellar populations, these galaxies will still emit a small amount of ultraviolet light. Each galaxy would be too faint to be detected individually by \textit{GALEX}, thus all of their light would be incorporated into the UV background. There may be a small additional contribution from the stars that make up the intracluster light. These stars are mostly stripped from cluster member galaxies \citep{demaio2018}, thus they would primarily be old, red stars.  However, there is some evidence of in-situ star formation in the intracluster medium \citep{Puchwein2010,Tonnesen2012}, which could enhance the UV contribution of intracluster stars. The sum of the unresolvable UV emission from these stars and the cluster galaxies could account for our measured excess. To test this hypothesis accounting for both cluster member galaxies and intracluster light, we used optical estimates of the cluster mass-to-light ratio, then calculated an UV surface brightness. 

We estimated the optical mass-to-light ratio using the best-fit relation in \cite{Popesso_MassLight}. This relation gives the luminosity in the SDSS r-band given a cluster mass. We then used 5 early-type SEDs (see Figure 4 of \cite{coe_relics}) exhibiting a range of UV upturn strengths (flux below 1600 \AA\ rest-frame) to calculate the luminosity in the \textit{GALEX} FUV bandpass. This FUV luminosity then allowed us to calculate the expected surface brightness of the unresolved cluster light.  For reference, Table \ref{table:massSB} shows a selection of surface brightness results for a range of cluster masses. We show both the high end surface brightness numbers, calculated assuming an SED with the strongest UV upturn and thus highest UV flux, and the low end surface brightness numbers calculated assuming the weakest UV upturn and lowest UV flux. Our final numbers quoted below are calculated from the median surface brightness of our entire cluster sample.

It is important to note that there is a sizeable uncertainty associated with the conversion between optical and ultraviolet luminosities. However, as a diagnostic tool to inform future studies, this uncertain calculation is useful.

\begin{table}
\centering
\begin{tabular}{ccc} 
\hline
Mass ($10^{14} M_{\odot}$) & S.B. High (CU) & S.B. Low (CU) \\
\hline
2.0 & 3.65 & 1.24 \\
4.0 & 4.82 & 1.64 \\ 
6.0 & 5.67 & 1.92 \\ 
8.0 & 6.36 & 2.16 \\ 
10.0 & 6.95 & 2.36 \\ 
\hline
\end{tabular} 
\caption{Surface brightnesses calculated from a range of cluster masses, as described in \ref{section:clusteremission}. Two surface brightnesses are quoted for each mass, one for the highest strength UV upturn and one for the lowest strength UV upturn.}
\label{table:massSB}
\end{table}

We applied our mass-to-light calculations to our full sample of \textit{Planck} clusters. To compare with our stacked analysis, we took the median of the calculated cluster UV brightnesses. This resulted in median UV brightness measurements of 1.6 CU to 4.7 CU, depending on the SED used for the luminosity conversion. While this is somewhat lower than our measured background excess, we believe it is likely a significant contributor.

\section{Conclusions}

Using \textit{GALEX} data, we measured a $12 \pm 2.3$ CU excess ($5.0\sigma$) of ultraviolet background light associated with $z\leq 0.3$ \textit{Planck} clusters at high Galactic latitudes ($ |b| > 60$). We conclude that the excess we measure is extragalactic in nature and directly related to the galaxy clusters. Based on our approximate calculations, we find the two most probable contributors to our measured UV excess are unresolved emission from quiescient cluster members, as well as emission from relativistic electrons in the intracluster medium. With the available data, we cannot accurately measure the relative contributions of each emission mechanism. 


Future work could improve upon this analysis by utilizing a more detailed background map. The improved resolution would create a more detailed radial profile which could be compared to cluster mass-to-light profiles and x-ray gas profiles to further test the relative contributions of each proposed emission mechanism. Enhanced resolution would also allow for the inclusion of higher redshift clusters. Including a larger cluster sample, for example an optically-selected sample with a greater mass and redshift range, could further improve constraints on the source of the UV excess.

Acknowledgements:

The authors thank David Schiminovich for conversations leading to the inception of this work, and Marc Postman for valuable insights throughout the process. We also thank the anonymous referee for their insightful comments. This work has been supported by NASA grant NNX17AC26G to the Johns Hopkins University. 

\bibliography{bib.bib}

\begin{thebibliography}{}
\expandafter\ifx\csname natexlab\endcsname\relax\def\natexlab#1{#1}\fi

\bibitem[{Akshaya {et~al.}(2018)Akshaya, Murthy, Ravichandran, Henry, \&
  Overduin}]{Akshaya_2018}
Akshaya, M.~S., Murthy, J., Ravichandran, S., Henry, R.~C., \& Overduin, J.
  2018, \apj, 858, 101

\bibitem[{{Akshaya} {et~al.}(2019){Akshaya}, {Murthy}, {Ravichandran}, {Henry},
  \& {Overduin}}]{Akshaya_2019}
{Akshaya}, M.~S., {Murthy}, J., {Ravichandran}, S., {Henry}, R.~C., \&
  {Overduin}, J. 2019, \mnras, 489, 1120

\bibitem[{{Becker} \& {Bolton}(2013)}]{becker}
{Becker}, G.~D., \& {Bolton}, J.~S. 2013, \mnras, 436, 1023

\bibitem[{{Bowyer}(1991)}]{bowyer91}
{Bowyer}, S. 1991, \araa, 29, 59

\bibitem[{{Chiang} {et~al.}(2019){Chiang}, {M{\'e}nard}, \&
  {Schiminovich}}]{chiang19}
{Chiang}, Y.-K., {M{\'e}nard}, B., \& {Schiminovich}, D. 2019, \apj, 877, 150

\bibitem[{{Coe} {et~al.}(2019){Coe}, {Salmon}, {Brada{\v{c}}}, {Bradley},
  {Sharon}, {Zitrin}, {Acebron}, {Cerny}, {Cibirka}, {Strait},
  {Paterno-Mahler}, {Mahler}, {Avila}, {Ogaz}, {Huang}, {Pelliccia}, {Stark},
  {Mainali}, {Oesch}, {Trenti}, {Carrasco}, {Dawson}, {Rodney}, {Strolger},
  {Riess}, {Jones}, {Frye}, {Czakon}, {Umetsu}, {Vulcani}, {Graur}, {Jha},
  {Graham}, {Molino}, {Nonino}, {Hjorth}, {Selsing}, {Christensen},
  {Kikuchihara}, {Ouchi}, {Oguri}, {Welch}, {Lemaux}, {Andrade-Santos}, {Hoag},
  {Johnson}, {Peterson}, {Past}, {Fox}, {Agulli}, {Livermore}, {Ryan}, {Lam},
  {Sendra-Server}, {Toft}, {Lovisari}, \& {Su}}]{coe_relics}
{Coe}, D., {Salmon}, B., {Brada{\v{c}}}, M., {et~al.} 2019, \apj, 884, 85

\bibitem[{{DeMaio} {et~al.}(2018){DeMaio}, {Gonzalez}, {Zabludoff}, {Zaritsky},
  {Connor}, {Donahue}, \& {Mulchaey}}]{demaio2018}
{DeMaio}, T., {Gonzalez}, A.~H., {Zabludoff}, A., {et~al.} 2018, \mnras, 474,
  3009

\bibitem[{{Hamden} {et~al.}(2013){Hamden}, {Schiminovich}, \&
  {Seibert}}]{hamden13}
{Hamden}, E.~T., {Schiminovich}, D., \& {Seibert}, M. 2013, \apj, 779, 180

\bibitem[{{Henry}(1991)}]{henry91}
{Henry}, R.~C. 1991, \araa, 29, 89

\bibitem[{{Hicks} {et~al.}(2010){Hicks}, {Mushotzky}, \&
  {Donahue}}]{hicks_coolcore}
{Hicks}, A.~K., {Mushotzky}, R., \& {Donahue}, M. 2010, \apj, 719, 1844

\bibitem[{{Inoue}(2014)}]{gammaray}
{Inoue}, Y. 2014, arXiv e-prints, arXiv:1412.3886

\bibitem[{{Martin} {et~al.}(2005){Martin}, {Fanson}, {Schiminovich},
  {Morrissey}, {Friedman}, {Barlow}, {Conrow}, {Grange}, {Jelinsky},
  {Milliard}, {Siegmund}, {Bianchi}, {Byun}, {Donas}, {Forster}, {Heckman},
  {Lee}, {Madore}, {Malina}, {Neff}, {Rich}, {Small}, {Surber}, {Szalay},
  {Welsh}, \& {Wyder}}]{martin05}
{Martin}, D.~C., {Fanson}, J., {Schiminovich}, D., {et~al.} 2005, \apjl, 619,
  L1

\bibitem[{{Morrissey} {et~al.}(2007){Morrissey}, {Conrow}, {Barlow}, {Small},
  {Seibert}, {Wyder}, {Budav{\'a}ri}, {Arnouts}, {Friedman}, {Forster},
  {Martin}, {Neff}, {Schiminovich}, {Bianchi}, {Donas}, {Heckman}, {Lee},
  {Madore}, {Milliard}, {Rich}, {Szalay}, {Welsh}, \& {Yi}}]{morrissey07}
{Morrissey}, P., {Conrow}, T., {Barlow}, T.~A., {et~al.} 2007, \apjs, 173, 682

\bibitem[{{Murthy}(2014)}]{murthy14}
{Murthy}, J. 2014, \apjs, 213, 32

\bibitem[{{Murthy}(2016)}]{murthy2016}
---. 2016, \mnras, 459, 1710

\bibitem[{{Murthy} {et~al.}(2010){Murthy}, {Henry}, \& {Sujatha}}]{murthy2010}
{Murthy}, J., {Henry}, R.~C., \& {Sujatha}, N.~V. 2010, \apj, 724, 1389

\bibitem[{{Penzias} \& {Wilson}(1965)}]{cmb}
{Penzias}, A.~A., \& {Wilson}, R.~W. 1965, \apj, 142, 419

\bibitem[{{Planck Collaboration} {et~al.}(2016){Planck Collaboration}, {Ade},
  {Aghanim}, {Arnaud}, {Ashdown}, {Aumont}, {Baccigalupi}, {Banday},
  {Barreiro}, {Barrena}, \& et~al.}]{psz}
{Planck Collaboration}, {Ade}, P.~A.~R., {Aghanim}, N., {et~al.} 2016, \aap,
  594, A27

\bibitem[{{Popesso} {et~al.}(2007){Popesso}, {Biviano}, {B{\"o}hringer}, \&
  {Romaniello}}]{Popesso_MassLight}
{Popesso}, P., {Biviano}, A., {B{\"o}hringer}, H., \& {Romaniello}, M. 2007,
  \aap, 464, 451

\bibitem[{{Puchwein} {et~al.}(2010){Puchwein}, {Springel}, {Sijacki}, \&
  {Dolag}}]{Puchwein2010}
{Puchwein}, E., {Springel}, V., {Sijacki}, D., \& {Dolag}, K. 2010, \mnras,
  406, 936

\bibitem[{{Sarazin}(1999)}]{Sarazin99}
{Sarazin}, C.~L. 1999, \apj, 520, 529

\bibitem[{{Sarazin}(2005)}]{sarazin_merger}
{Sarazin}, C.~L. 2005, in X-Ray and Radio Connections, ed. L.~O. {Sjouwerman}
  \& K.~K. {Dyer}, 8.01

\bibitem[{{Sunyaev} \& {Zeldovich}(1970)}]{sz_effect}
{Sunyaev}, R.~A., \& {Zeldovich}, Y.~B. 1970, \apss, 7, 3

\bibitem[{{Tonnesen} \& {Bryan}(2012)}]{Tonnesen2012}
{Tonnesen}, S., \& {Bryan}, G.~L. 2012, \mnras, 422, 1609

\bibitem[{{Upton Sanderbeck} {et~al.}(2018){Upton Sanderbeck}, {McQuinn},
  {D'Aloisio}, \& {Werk}}]{upton-sanderbeck}
{Upton Sanderbeck}, P.~R., {McQuinn}, M., {D'Aloisio}, A., \& {Werk}, J.~K.
  2018, \apj, 869, 159

\bibitem[{{Voyer} {et~al.}(2011){Voyer}, {Gardner}, {Teplitz}, {Siana}, \& {de
  Mello}}]{Voyer11}
{Voyer}, E.~N., {Gardner}, J.~P., {Teplitz}, H.~I., {Siana}, B.~D., \& {de
  Mello}, D.~F. 2011, \apj, 736, 80

\bibitem[{{Xu} {et~al.}(2005){Xu}, {Donas}, {Arnouts}, {Wyder}, {Seibert},
  {Iglesias-P{\'a}ramo}, {Blaizot}, {Small}, {Milliard}, {Schiminovich},
  {Martin}, {Barlow}, {Bianchi}, {Byun}, {Forster}, {Friedman}, {Heckman},
  {Jelinsky}, {Lee}, {Madore}, {Malina}, {Morrissey}, {Neff}, {Rich},
  {Siegmund}, {Szalay}, \& {Welsh}}]{xu05}
{Xu}, C.~K., {Donas}, J., {Arnouts}, S., {et~al.} 2005, \apj, 619, L11

\end{thebibliography}

\appendix 

\section{List of Clusters}

\begin{longtable}{ccc}

\hline
Name & Redshift & SZ Mass ($10^{14} M_{\odot}$) \\
\hline

ACO S 1109 & 0.140 & 2.8 \\ 
RMJ140358.9+154409.6 & 0.181 & 3.5 \\ 
RMJ143312.4+122756.8 & 0.236 & 4.8 \\ 
ACO S 1077 & 0.312 & 7.7 \\ 
RXCJ0014.3-3023 & 0.307 & 9.8 \\ 
RXCJ0011.3-2851 & 0.062 & 2.6 \\ 
RXC J1341.8+2622 & 0.072 & 2.7 \\ 
RXC J1348.8+2635 & 0.062 & 4.5 \\ 
RXCJ2351.6-2605 & 0.226 & 7.6 \\ 
RXC J2326.2-2406 & 0.088 & 2.2 \\ 
RMJ141438.9+270311.1 & 0.481 & 7.1 \\ 
ABELL 2663 & 0.244 & 3.9 \\ 
RMJ140649.4+274556.8 & 0.566 & 5.7 \\ 
RXC J1359.2+2758 & 0.061 & 2.6 \\ 
RXC J1349.3+2806 & 0.075 & 2.6 \\ 
RMJ145332.1+280358.0 & 0.257 & 4.3 \\ 
RXCJ0020.7-2542 & 0.141 & 4.3 \\ 
ACO 2538 & 0.083 & 2.6 \\ 
PSZ2 G048.21-65.00 & 0.420 & 5.6 \\ 
A1961 & 0.234 & 4.7 \\ 
PSZ2 G056.62+88.42 & 0.045 & 3.3 \\ 
RXC J1259.7+2756 & 0.023 & 7.2 \\ 
RMJ144415.9+355713.1 & 0.361 & 5.0 \\ 
PSZ2 G061.75+88.11 & 0.044 & 3.4 \\ 
RXCJ2325.3-1207 & 0.085 & 2.5 \\ 
GMBCG J204.74580+32.97396 & 0.273 & 4.1 \\ 
RMJ142140.1+371728.7 & 0.163 & 3.8 \\ 
RXC J1322.8+3138 & 0.308 & 6.6 \\ 
WHL J215.168+39.91 & 0.609 & 6.7 \\ 
RMJ140344.1+382703.9 & 0.485 & 5.8 \\ 
RXC J2341.2-0901 & 0.251 & 4.1 \\ 
RMJ142716.1+440730.6 & 0.502 & 5.2 \\ 
RXCJ2354.2-1024 & 0.076 & 2.8 \\ 
RMJ140026.7+410140.5 & 0.250 & 4.2 \\ 
WHL J357.962-8.991 & 0.394 & 5.9 \\ 
RXC J1413.7+4339 & 0.089 & 2.1 \\ 
RXC J1305.9+3054 & 0.183 & 4.7 \\ 
RXCJ2344.2-0422 & 0.079 & 3.1 \\ 
ZwCl 1341.2+4022 & 0.222 & 3.8 \\ 
RMJ234517.1-030238.9 & 0.355 & 4.4 \\ 
PSZ2 G089.39+69.36 & 0.680 & 5.7 \\ 
RMJ142210.2+483414.7 & 0.070 & 1.8 \\ 
RXC J0003.1-0605 & 0.232 & 6.0 \\ 
RXC J1335.3+4059 & 0.228 & 8.1 \\ 
RXC J1351.7+4622 & 0.062 & 1.5 \\ 
RXCJ0043.4-2037 & 0.292 & 8.1 \\ 
RXC J1332.7+5032 & 0.280 & 8.2 \\ 
RMJ003353.1-075210.4 & 0.304 & 5.6 \\ 
RXC J1313.1+4616 & 0.183 & 4.3 \\ 
RXCJ0034.6-0208 & 0.081 & 2.3 \\ 
RXC J1315.1+5149 & 0.284 & 6.8 \\ 
RXC J1306.9+4633 & 0.226 & 5.7 \\ 
RXCJ0041.8-0918 & 0.056 & 4.9 \\ 
RMJ004330.7-101009.1 & 0.502 & 5.9 \\ 
RMJ130122.0+481545.0 & 0.255 & 3.8 \\ 
RMJ004512.5-015231.7 & 0.557 & 6.4 \\ 
RMJ124930.9+494902.3 & 0.284 & 4.4 \\ 
RXCJ0056.3-0112 & 0.044 & 3.4 \\ 
ACO  117 & 0.054 & 1.5 \\ 
RMJ005815.8-065213.7 & 0.119 & 4.1 \\ 
ZwCl 0102.4-0012 & 0.277 & 5.7 \\ 
RXC J1229.0+4737 & 0.254 & 5.9 \\ 
RMJ121912.2+505435.3 & 0.544 & 6.0 \\ 
RXCJ0115.2+0019 & 0.045 & 1.5 \\ 
RXCJ0108.8-1524 & 0.053 & 1.8 \\ 
RMJ115914.9+494748.4 & 0.363 & 7.6 \\ 
RXC J1149.0+5135 & 0.132 & 2.7 \\ 
RXC J1218.4+4013 & 0.304 & 4.8 \\ 
RXCJ0120.9-1351 & 0.052 & 1.7 \\ 
WHL J24.3324-8.477 & 0.566 & 8.9 \\ 
RMJ115353.2+425213.2 & 0.333 & 4.4 \\ 
PSZ2 G155.95-72.13 & 0.620 & 6.5 \\ 
RMJ121731.2+364111.3 & 0.377 & 5.6 \\ 
RXCJ0131.8-1336 & 0.206 & 8.5 \\ 
RXC J0137.9-1248 & 0.211 & 4.6 \\ 
WHL J170.907+43.05 & 0.196 & 3.7 \\ 
RXC J0159.8-0850 & 0.405 & 7.2 \\ 
ACO 1319 & 0.288 & 4.8 \\ 
ACO 1401 & 0.165 & 3.1 \\ 
RXC J1111.6+4050 & 0.079 & 2.5 \\ 
RXCJ0105.5-2439 & 0.230 & 5.7 \\ 
RXCJ0206.4-1453 & 0.153 & 4.0 \\ 
RMJ105038.6+354912.4 & 0.509 & 5.7 \\ 
ACO  2985 & 0.174 & 3.3 \\ 
RXCJ0236.6-1923 & 0.091 & 2.5 \\ 
RMJ110444.3+283541.3 & 0.580 & 5.9 \\ 
PSZ2 G205.05-62.95 & 0.310 & 7.6 \\ 
WHL J174.518+27.97 & 0.447 & 7.4 \\ 
RXC J1212.3+2733 & 0.353 & 7.5 \\ 
PSZ2 G208.61-74.39 & 0.720 & 6.3 \\ 
PSZ2 G213.27+78.38 & 0.316 & 4.8 \\ 
RXC J1129.8+2347 & 0.137 & 3.6 \\ 
RXC J0241.3-2839 & 0.232 & 6.2 \\ 
RXC J0227.2-2851 & 0.214 & 4.1 \\ 
RXC J1123.9+2129 & 0.190 & 5.3 \\ 
RXC J0152.5-2853 & 0.413 & 5.8 \\ 
RXC J1155.3+2324 & 0.143 & 5.9 \\ 
RXC J1123.2+1935 & 0.104 & 2.9 \\ 
RMJ113608.5+201913.0 & 0.321 & 5.4 \\ 
RXC J1113.3+1735 & 0.171 & 3.3 \\ 
A1367 & 0.021 & 1.7 \\ 
RXC J1112.9+1326 & 0.169 & 4.5 \\ 
RXC J0159.0-3412 & 0.410 & 8.1 \\ 
RXC J1132.8+1428 & 0.083 & 3.9 \\ 
RXCJ0220.9-3829 & 0.228 & 3.8 \\ 
RXCJ0225.9-4154 & 0.220 & 6.1 \\ 
RXCJ0232.2-4420 & 0.284 & 7.5 \\ 
ACO  3036 & 0.190 & 3.8 \\ 
SPT-CLJ0218-4315 & 0.560 & 5.1 \\ 
RXC J0212.8-4707 & 0.115 & 2.3 \\ 
RMJ120709.6+092344.2 & 0.292 & 4.5 \\ 
SPT-CLJ0150-4511 & 0.320 & 3.6 \\ 
NSCS J123631+190301 & 0.180 & 3.4 \\ 
RXC J1229.9+1147 & 0.085 & 3.0 \\ 
RXC J1227.4+0849 & 0.090 & 2.2 \\ 
RXC J1230.7+1033 & 0.165 & 5.6 \\ 
SPT-CLJ0114-4123 & 0.213 & 4.5 \\ 
RMJ122241.3+020558.9 & 0.229 & 4.1 \\ 
RXC J1241.3+1834 & 0.073 & 2.6 \\ 
RXC J1234.2+0947 & 0.229 & 5.9 \\ 
RXCJ0110.0-4555 & 0.024 & 1.1 \\ 
RXC J0051.1-4833 & 0.187 & 3.2 \\ 
RXCJ1258.6-0145 & 0.084 & 4.4 \\ 
SPT-CLJ0040-4407 & 0.350 & 5.8 \\ 
RXC J1311.5-0120 & 0.183 & 8.8 \\ 
WHL J197.343+10.85 & 0.426 & 6.1 \\ 
ZwCl 1324.6+0229 & 0.259 & 6.4 \\ 
RXC J1303.7+1916 & 0.064 & 1.9 \\ 
RMJ133111.4+010003.2 & 0.359 & 5.6 \\ 
RXCJ1342.0+0213 & 0.076 & 2.5 \\ 
RMJ134431.8+022244.3 & 0.376 & 5.9 \\ 
RXC J1353.0+0509 & 0.079 & 2.5 \\ 
SPT-CLJ2344-4243 & 0.596 & 8.9 \\ 
RXCJ0025.5-3302 & 0.049 & 1.5 \\ 
RXCJ2313.9-4244 & 0.056 & 1.5 \\ 
ACO S 1121 & 0.358 & 7.2 \\ 
RMJ135620.4+104724.2 & 0.247 & 4.3 \\ 
RXCJ0006.0-3443 & 0.115 & 4.0 \\ 
RXC J1354.0+1455 & 0.125 & 2.9 \\ 
RXCJ2357.0-3445 & 0.047 & 2.5 \\ 
ZwCl 1357.4+1430 & 0.209 & 5.7 \\ 
RXCJ2315.7-3746 & 0.179 & 5.2 \\  

\end{longtable}

\end{document}